\documentclass{kapproc} % Computer Modern font calls

\RequirePackage{graphicx}%
\RequirePackage{epsf}%
\input{psfig.sty}

\upperandlowercase
\let\footnote\savefootnote
\let\footnotetext\savefootnotetext 
 
\kluwerbib % will produce  bibliography 
\begin{document}

%------------ article title  ------------------->>

\articletitle{ Monte-Carlo experiments on star-cluster
      induced integrated-galaxy IMF variations}

%% optional, to supply a subtitle:
%\articlesubtitle{Spineto@50}

%% Supply a shorter version of the title for the running head:
\chaptitlerunninghead{Monte-Carlo experiments on IMF variations}

%------ author/affiliation choices -------------->>

%% Single author or several authors with same affiliation

 \author{Carsten Weidner and Pavel Kroupa}
 \affil{Observatory of the University of Bonn, Auf dem H\"ugel 71, D-53121
    Bonn, Germany}
 \email{cweidner@astro.uni-bonn.de, pavel@astro.uni-bonn.de}

%% Multiple authors, multiple affiliations

%\author{First Author\altaffilmark{1}, Second Author\altaffilmark{2}, 
%         Third Author\altaffilmark{1,3}}

%\affil{\altaffilmark{1}Institute, Address, Country, \\ 
%\altaffilmark{2}Institute, Address, Country, \\
%\altaffilmark{3}Institute, Address, Country}

%\email{author1@add1,author2@add2,author3@add3}

% abstract
 \begin{abstract}
As most if not all stars are born in stellar clusters the shape of the
mass function of the field stars is not only determined by the initial
mass function of stars (IMF) but also by the cluster mass function
(CMF). In order to quantify this Monte-Carlo simulations were carried
out by taking cluster masses randomly from a CMF and then populating
these clusters with stars randomly taken from an IMF. Two cases were
studied. Firstly the star masses were added randomly until the cluster
mass was reached. Secondly a number of stars, given by the cluster mass
divided by an estimate of the mean stellar mass and sorted by mass,
were added until the desired cluster mass was reached. Both
experiments verified the analytical results of Kroupa \& Weidner (2003)
that the resulting integrated stellar initial mass function is a
folding of the IMF with the CMF and 
therefore steeper than the input IMF above 1 $M_{\odot}$. 

 \end{abstract}
%------------ body of article ------------------->>
\section{The Integrated Galactic Initial Mass
Function from Clustered Star Formation} 
\label{sec:meth}
\noindent
Kroupa \& Weidner (2003) showed that the integrated galactic stellar
initial mass 
function (IGIMF) is obtained by summing up the 
stellar IMFs contributed by all the star clusters that formed over the
age of a galaxy. In their approach the mass of the most massive
star in an embedded cluster with stellar mass $M_{\rm ecl}$ is
calculated from by
$
1 = \int_{m_{\rm max}}^{m_{\rm max*}} \xi(m)\,dm~{\rm and}~
%\label{eq:mm}
M_{\rm ecl} = \int_{m_{\rm l}}^{m_{\rm max}} m\,\xi(m)\,dm.
$
The resulting function $m_{\rm
max} = {\rm fn}(M_{\rm ecl})$ is quantified by \cite{WeKr04} who infer
that there exists a fundamental 
upper stellar mass limit, $m_{\rm max*}\approx150\,M_\odot$, because
otherwise the populous cluster R136 would contain too many stars with
m $> 100 M_{\odot}$.
\section{Results from the Monte-Carlo simulations}
\label{sec:montec}
In order to verify the above results we here perform Monte-Carlo
simulations of 
star clusters and the stars within. 50 Million clusters are randomly
chosen from a power-law CMF between 5 $M_{\odot}$ and
$10^{6}\, M_{\odot}$ and with a Salpeter Index $\beta=2.35$. These clusters are
then filled with stars in two ways. First stars were taken randomly
from a canonical Kroupa-IMF and their masses added until the sum equalled
the chosen 
cluster ({\it long dashed line} in Fig.~1). In the second
way, a number of stars given by the mass 
of the cluster divided by an estimate of the mean stellar mass are
randomly chosen, sorted and then added - starting with the least
massive star ({\it short dashed line} in Fig.~1). The addition is
terminated as soon as the cluster mass is reached, or the process is
repeated if the cluster mass is not reached. The sorted IGIMF agrees
very well with the 
semi-analytical results while the IGIMF from pure random adding shows
a somewhat less steeper slope. 

\begin{minipage}[t]{5cm}
\psfig{file=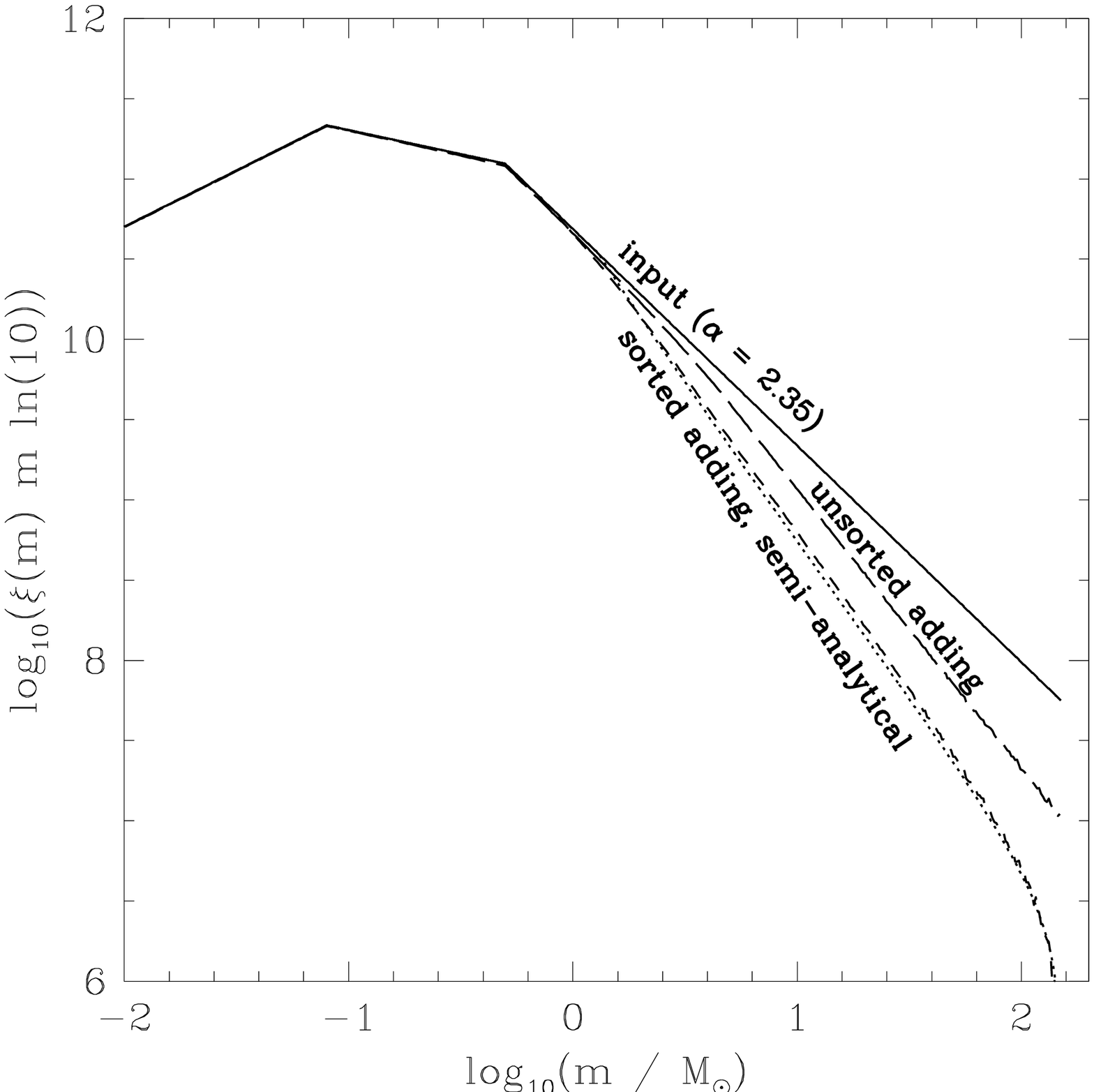,width=5.0cm}
\end{minipage}
\begin{minipage}[t]{5cm}
\vspace*{-4.5cm}
{{\it Figure 1}: Solid line: Canonical stellar IMF with
  $\alpha=2.35$ for $m>1\,M_\odot$. Dotted line: 
  IGIMF resulting from the semi-analytical approach with $\beta=2.35$. 
Short dashed line: IGIMF gained after sorted adding of
random stars. Long dashed line: IGIMF produced by
unsorted adding of stars.}
\end{minipage}

\section{Discussion and Conclusions}
\label{sec:concs}
\begin{itemize}
\item[-] The Monte-Carlo simulations confirm the results of our
  semi-analytical formalism.
\item[-] integrated galactic IMFs must always be steeper for $m>1\,M_\odot$
than the stellar IMF that results from a local star-formation
event. 
\end{itemize}

\end{document}